\documentclass{aa}
\usepackage[utf8]{inputenc}

\usepackage{txfonts}
\usepackage{graphicx}
\usepackage{epstopdf}
\usepackage{natbib} 
\bibpunct{(}{)}{;}{a}{}{,} 
\title{Vortex Flows in the Solar Chromosphere}
\subtitle{I. Automatic detection method}
\author{Y. Kato\inst{1}
        \and
        S. Wedemeyer\inst{1}
        }
 \institute{Institute of Theoretical Astrophysics, University of Oslo, P.O. Box 1029 Blindern, N-0315 Oslo, Norway\\
              \email{yoshiaki.kato@astro.uio.no, sven.wedemeyer@astro.uio.no}
             }
\date{}

\abstract{
   Solar ``magnetic tornadoes'' are produced by rotating magnetic field structures that extend from the upper convection zone and the photosphere to the corona of the Sun.  
   Recent studies show that such rotating features are an integral part of atmospheric dynamics and occur on a large range of spatial scales. 
   A systematic statistical study of magnetic tornadoes is a necessary next step towards understanding their formation and their role for the mass and energy transport in the solar atmosphere.
   For this purpose, we have developed a new automatic detection method for chromospheric swirls, i.e. the observable signature of solar tornadoes or, more generally, chromospheric vortex flows and rotating motions.  
  Unlike the previous studies that relied on visual inspections, our new method combines a line integral convolution (LIC) imaging technique and a scalar quantity which represents a vortex flow on a two-dimensional plane.
  We have tested two detection algorithms, based on the enhanced vorticity and vorticity strength quantities, by applying them to 3D numerical simulations of the solar atmosphere with CO5BOLD.
   We conclude that the vorticity strength method is superior compared to the enhanced vorticity method in all aspects. 
   Applying the method to a numerical simulation of the solar atmosphere revealed very abundant small-scale, short-lived chromospheric vortex flows that had not been found by visual inspection before. }


\begin{document}

\maketitle

\section{Introduction}
\label{sec:intro}

A solar ``magnetic tornado'' is generated when the footpoint of a magnetic field structure coincides with an intergranular vortex flow in the photosphere and topmost parts of the convection zone \citep{wedemeyer+steiner:2014}. 
The rotation is mediated by the magnetic field into the chromosphere, where the plasma is forced to follow the rotating magnetic field and thus creates an observable vortex flow, which is referred to as ``chromospheric swirl'' \citep{wedemeyer-bohm+rouppe_van_der_voort:2009}. 
A small-scale quiet Sun vortex was also observed by \citet{2016A&A...586A..25P} in different chromospheric spectral lines. 
Solar tornadoes extend further into the corona above and thus provide channels for the transfer of magneto-convective mechanical energy into the upper solar atmosphere, thus making them a potential candidate for contributing to the heating of the chromosphere and corona. 
\citet{wedemeyer-bohm+:2012} found that, based on numerical simulations, a sufficient amount of Poynting flux could be carried upwards in a magnetic tornado. 
A more detailed determination of the net energy flux, which is in principle provided in the upper solar atmosphere, requires a systematic statistical analysis of the properties of solar tornadoes.  
Previous studies relied on visual inspections of chromospheric observations, mostly in the Ca\,II infrared triplet line at 854.2\,nm, and produced rather small sample sizes.  
For instance, \citet{wedemeyer-bohm+:2012} detected 14 swirls within a field of view of  55"\,$\times$\,55" during on observation with a duration of 55\,min, implying a occurrence of $1.9\times 10^{-4}$~vortices Mm$^{-2}$ min$^{-1}$ and an average lifetime of $(12.7 \pm 4.0)$\,min. 
The occurrence rate of chromospheric vortex flows seems to be roughly one order magnitude smaller than the corresponding rate for photospheric vortices of $3.1\times 10^{-3}$~vortices Mm$^{-2}$ min$^{-1}$ found by \citet{bonet+:2010}. 
A lower occurrence of chromospheric vortices is expected to some extent when assuming that the formation of a chromospheric vortex requires both a driving photospheric vortex and a co-located magnetic field structure \citep{wedemeyer+steiner:2014}. 
However, given the challenges with reliably spotting chromospheric swirls as part of the complex dynamics exhibited in such observational image sequences, the derived occurrence rate has to be considered a rough estimate and most likely a lower limit only. 
Apart from the unknown effect of yet undetected vortex flows on the transport of energy and  mass, more accurate occurrence rates of vortex flows could help to characterise the nature of turbulent flows in the solar atmosphere.

In this first paper of a series, we describe an automatic detection method that is capable of identifying the observable signatures of magnetic tornadoes in complex flow patterns, which is a necessary next step towards a quantitative determination of the possible contribution of magnetic tornadoes to the heating of the upper solar atmosphere.  
The method was developed and tested with numerical simulations produced with CO\raisebox{0.5ex}{\footnotesize 5}BOLD \citep{freytag+:2012}. 
After the description of the method in Sect.~\ref{sec:method}, which combines line integral convolution (LIC) imaging techniques, vorticity, and vorticity strength, we demonstrate the results for one of the CO\raisebox{0.5ex}{\footnotesize 5}BOLD simulations in Sect.~\ref{sec:results}. 
After a discussion of the results in Sect.~\ref{sec:discussion}, 
conclusions and an outlook are provided in Sect.~\ref{sec:conclusions}.
In forthcoming papers of this series, the full analysis results for a sequence of numerical simulations and for observational data sets will be presented. 

\section{Method}
\label{sec:method}

%
The method presented here consists of the following steps: 
\begin{enumerate}
\item Determination of the velocity field (Sect.~\ref{subsec:velocity_field}).
\item Vortex detection (Sect.~\ref{subsec:vortex_detection}) based on  
vorticity and vorticity strength (Sect.~\ref{subsec:vorticity_vorticity-strength}). 
\item Event identification (Sect.~\ref{subsec:event_identification}). 
\end{enumerate}
In Appendix~\ref{sec:appendix1}, we show the flowchart of our method step by step.
The method is tested and illustrated first with a very simplified vortex flow (see Sect.~\ref{subsec:model}) and then with a more realistic combination of a numerical model atmospheres and a fixed vortex flow (see Sect.~\ref{subsec:marker_tests}). 
%

\subsection{Determination of the velocity field}
\label{subsec:velocity_field}

%
Quantitative studies of atmospheric dynamics \citep[e.g.,][]{november:1986, november+simon:1988} require that the velocity field in the solar atmosphere is known with sufficient accuracy. 
Unfortunately, the pronounced dynamics, which occurs on a large range of temporal and spatial scales, make the reliable determination of the atmospheric velocity field from observations a challenging task. 
Next to the complexity of (multidimensional) flows in the solar atmosphere itself, observational effects, such as intermediate blurring as a result of varying seeing condition, and limited spatial resolution make it difficult to find swirling features.  
Some of the difficulties can be overcome by advanced feature tracking techniques such as the commonly known local correlation tracking (LCT). 
For instance, the FLCT method by \citet{welsch+:2004} and \citet{fisher+welsch:2008} is able to determine the velocity field from sequences of observational images with sufficient quality.

It has nevertheless to be checked thoroughly  how well the derived LCT velocities match the actual physical velocities. 
Such tests can be performed on basis of numerical simulations for which the initial physical velocities are known and can be compared to corresponding LCT velocites derived from synthetic observations of the simulated atmospheric dynamics. 
In this first paper of a series, however, we present a precursory step for  which we use one of the model atmospheres created with  CO\raisebox{0.5ex}{\footnotesize 5}BOLD (see  Sect.~\ref{subsec:model}) for testing our detection methods as described below.
In future parts of this series, velocities derived with the FLCT method will be used and the implications of the inherently  
limited velocity accuracy on the detection of vortex flows will be discussed.

\begin{figure*}
\centering
\includegraphics[width=1.0\hsize]{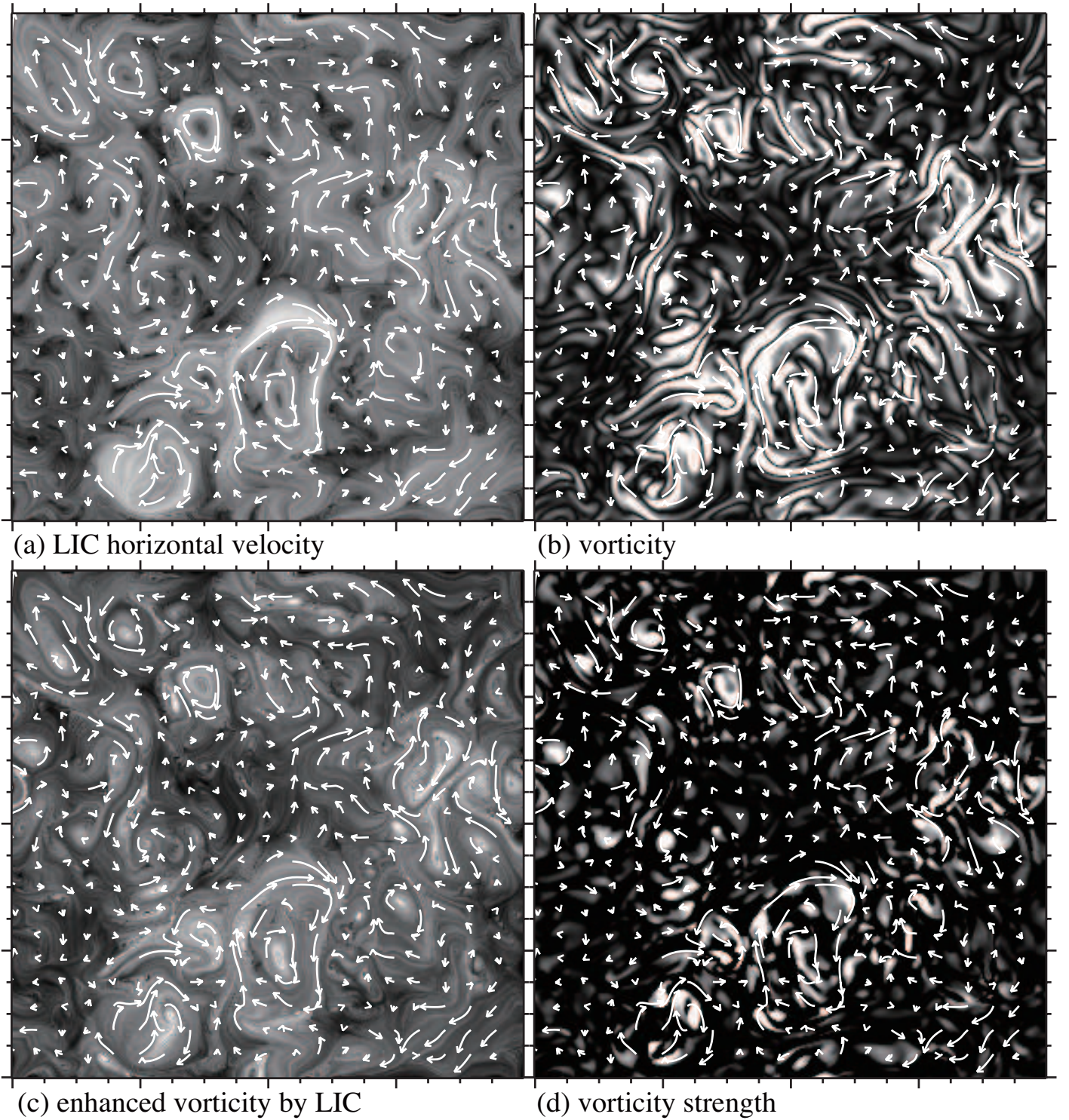}
\caption{Sample images for illustrating velocity distribution of a snapshot in our simulation ($8\,{\rm Mm}\times 8\,{\rm Mm}$). (a) LIC imaging scaled by the horizontal velocity amplitude, (b) vorticity, (c) LIC imaging scaled by the vorticity amplitude, and (d) vorticity strength.  Arrows indicate the velocity field.}
\label{fig:images}%
\end{figure*}

\begin{figure*}[th]
\sidecaption
\includegraphics[width=12cm]{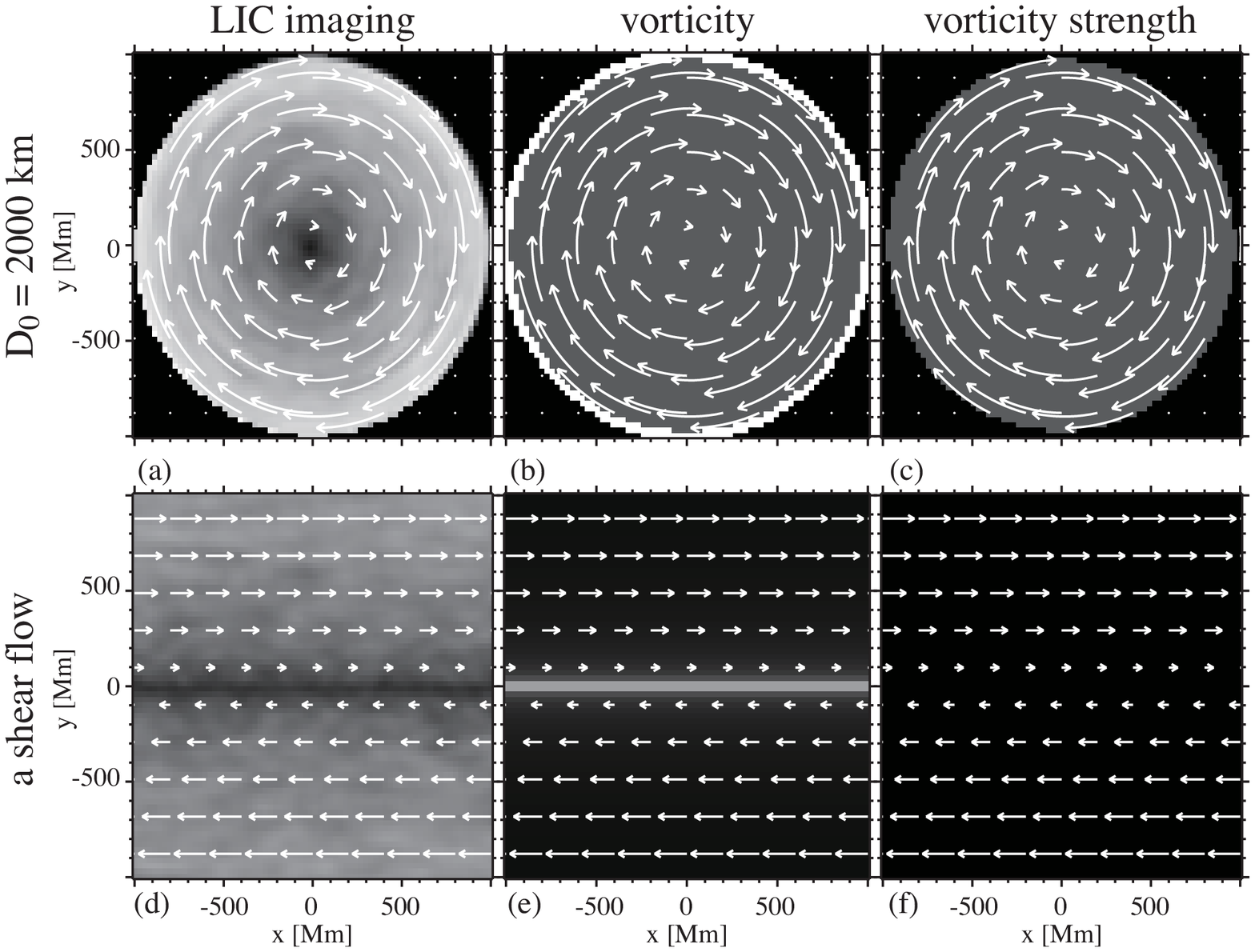}
\caption{Comparison between LIC imaging, vorticity, and vorticity strength of a solid rotation (upper panels) and that of a shear flow (lower panels) indicated by white arrows.  Upper panels: (a) the horizontal velocity amplitude of a swirling feature is illustrated by the LIC imaging, (b) the tangential discontinuity at the edge of a swirling feature is enhanced by the vorticity, and (c) the constant angular rotation speed of a swirling feature is depicted by the vorticity strength.  Lower panels: (d) same as (a) but for a shear flow, (e) a gap between the rightward and leftward velocity at $y=0$ is illustrated by the vorticity, and (f) there are no features shown by the vorticity strength.}
\label{fig:sample}%
\end{figure*}

\subsection{Model atmosphere for testing}
\label{subsec:model}

The model was computed with the 3D radiation magnetohydrodynamic code CO\raisebox{0.5ex}{\footnotesize 5}BOLD \citep{freytag+:2012} and is equivalent to the one used by \citet{wedemeyer-bohm+:2012} but it was produced with an improved numerical integration scheme \citep{steiner+:2013}, resulting in reduced numerical diffusivity and thus more structure on smaller spatial scales.
The new model was constructed by superimposing an initial magnetic field on a snapshot of a hydrodynamic simulation and advanced in time until it had relaxed from this initial condition. 
The initial magnetic field is strictly vertical and has a field strength of $|B_0| = 50$\,G. 

The computational box comprises $286\,\times\,286\,\times\,286$ grid cells and has a horizontal extent of $8.0\,\mathrm{Mm}\,\times\,8.0\,\mathrm{Mm}$ with a constant grid cell width of $\Delta x = \Delta y = 28$\,km. 
The grid is non-equidistant in height and extents from -2.4\,Mm at the bottom (in the convection zone) to 2.0\,Mm at the top (in the model chromosphere) with a constant vertical grid spacing of $\Delta z = 12$\,km at all $z > -128$\,km and gradually increasing to $\Delta z = 28.2$\,km for all $z < -1132$\,km. 

The computational time steps are typically on the order of a few milliseconds. 
A sequence with a duration of 60\,min and a cadence of 1\,s is produced of which the first 30\,min are used for the tests presented in this paper. 
%

\subsection{Identification of vortices}
\label{subsec:vorticity_vorticity-strength}

An objective definition of a vortex has yet to be established and therefore the identification of vortices in complex/turbulent flows has been a primary subject \citep[see, e.g.,][and references therein]{1995JFM...285...69J}.
The so-called line integral convolution (LIC) technique turns out to be helpful in this respect. 
The LIC techniques allows to visualise turbulent flows and was first introduced by \citet{cabral+:1993}.  
A LIC image for a given velocity field is created by tracing streamlines whose intensity is proportional to the horizontal velocity amplitude.  
This technique can enhance the flow pattern of small-scale and large-scale eddies simultaneously, as can be seen in Fig.~\ref{fig:images}\,a, and therefore it can help to identify vortices by visual and automatic inspection.
However, an automatic detection algorithm for vortices requires  quantities that enables the objective, robust and reproducible identification of vortices. 

A vortex flow is by definition associated with a high value of the \emph{vorticity} $\vec{\omega}$, which is defined as  
\begin{equation}
\vec{\omega} = \vec{\nabla} \times \vec{v}\,.
\label{eqn:vorticity}
\end{equation}
Unfortunately, the definition of the vorticity in \mbox{Eq.~(\ref{eqn:vorticity})} also results in high values for shear flows, i.e., flows with opposite direction, which thus have to be distinguished from the sought after true circular vortex flows.
This ambiguity can be avoided by using the \emph{vorticity strength} instead, which is the imaginary part of the eigenvalues of the velocity gradient tensor. 
The vorticity strength has indeed been successfully used by \citet{moll+:2011} for investigating flow patterns in the surface-near layers of the convection zone and the photosphere.   
In our study, we use the same technique to investigate flow patterns in the chromosphere. 
In the following, we briefly demonstrate how the vorticity strength is used for finding swirling features in flows projected on a \emph{two-dimensional} view plane.
Consider the velocity gradient tensor, $\mathcal{D}_{ij}$
\begin{equation}
 \mathcal{D}_{ij}=\frac{\partial v_{\rm i}}{\partial x_{\rm j}}
\label{eqn:Dij}
\end{equation}

If $\lambda$ are the eigenvalues of $\mathcal{D}_{ij}$, then
\begin{equation}
\mathcal{D}_{ij} - \lambda I=0
\label{eqn:tensor_equation}
\end{equation}
where $I$ is the second order unit tensor.

The eigenvalues can be determined by solving the characteristic equation
\begin{equation}
\det{\left[\mathcal{D}_{ij} - \lambda  I\right]} =0 
\label{eqn:determinant_tensor}
\end{equation}
which, for a velocity flow in two-dimensional space \mbox{$\vec{v} = (v_\mathrm{x},v_\mathrm{y})$}, can be written as
\begin{equation}
\lambda^2 + P\lambda + Q = 0
\label{eqn:determinant_tensor_equation}
\end{equation}
where $P = -\mathrm{tr}{(\mathcal{D}_{ij})}$ and $Q=\det{(\mathcal{D}_{ij})}$. 
Equation\,(\ref{eqn:determinant_tensor_equation}) has the following canonical solutions: 
\begin{equation}
\lambda = \frac{-P\pm\sqrt{P^2 - 4Q}}{2}.
\label{eqn:canonical_solution}
\end{equation}
These solutions contain (i)~one real root, (ii)~two distinctive real roots, or (iii)~a conjugate pair of complex roots.  
In general, positive and negative signs of the real part of the solutions, $\operatorname{Re}(\lambda)$, indicate diverging and converging flows, respectively. 
The imaginary part of the solutions, $\operatorname{Im}(\lambda)$, indicates the strength of the angular rotation speed of a circular motion.  
For a complete description in terms of eigenvalues and eigenvectors of the velocity gradient tensor, see \citet{chong+:1990} for all possible cases of flow field geometries in three-dimensional space.

As an example, we consider a circular flow centered at the origin, $(x,y) = (0,0)$, with two components: (i)~A constant angular rotation frequency $\nu$ in the counter-clockwise direction and (ii)~a radial flow with a constant velocity $\mu$ directed outwards away from the origin.  
First, the velocity is transformed from polar coordinates to Cartesian coordinates: 
\begin{equation}
\left[\begin{array}{c}  v_{\rm x} \cr  
                                  v_{\rm y}
\end{array}\right]=
\left[\begin{array}{lr} \cos{\theta} & -\sin{\theta} \cr  
                                   \sin{\theta} & \cos{\theta}
\end{array}\right]
\left[\begin{array}{c}  v_{\rm r} \cr  
                                  v_{\rm \theta}
\end{array}\right]
\end{equation}
where $v_{\rm r} = \mu$ and $v_{\theta}=r\nu$ and $r=\sqrt{x^{2} + y^{2}}$ is the distance from the origin.  The velocity vector can be represented as
\begin{equation}
\mbox{\boldmath$v$}=\left[\mu\left(\frac{x}{r}\right) -\nu y\right]\mbox{\boldmath$e$}_{\rm x} + \left[\mu\left(\frac{y}{r}\right) + \nu x\right]\mbox{\boldmath$e$}_{\rm y}
\end{equation}

The velocity gradient tensor of this case can be represented as
\begin{equation}
\mathcal{D}_{\rm ij}=\left[\begin{array}{lr} \mu\left(\frac{x}{r^2}\right) & -\nu \cr  
                                                 \nu & \mu\left(\frac{y}{r^2}\right)
\end{array}\right], 
\end{equation}
which results in 
\begin{equation}
P=\frac{(x + y)}{r^{2}}\mu
\end{equation}
and\begin{equation}
Q=\left(\frac{xy}{r^{4}}\right)\mu^{2}+\nu^{2}
\end{equation}
(see Eq.~(\ref{eqn:determinant_tensor_equation})). 
For a pure circular flow with $\mu=0$ and $\nu\neq 0$, the solution consists of a conjugate pair of complex roots of $\lambda=\pm i\nu$. 
In contrast, for a radial, non-circular flow with $\mu\neq 0$ and $\nu=0$, the solution has two distintive real roots of $\lambda=\mu/r^{2}x$ and $\mu/r^{2}y$.

\subsection{\emph{Vortex detection}}
\label{subsec:vortex_detection}

Once the velocity field is provided, the vorticity and the vorticity strength are calculated according to Eqs.\,(\ref{eqn:vorticity})-(\ref{eqn:canonical_solution}). %
The vorticity  in a chromospheric layer in the simulations exhibits a complicated filamentary structure as shown in Fig.~\ref{fig:images}\,b.
Prior to the detection procedure, the LIC technique is applied to the resulting sequence of vorticity maps in order to smooth and to enhance not only eddies but also circular flows so that they become easier to detect  (see Fig.~\ref{fig:images}c). 
This preparatory step has proven to reduce the computational costs of the vortex detection procedure.
In contrast, the corresponding vorticity strength map is composed of more isolated features as shown in Fig.~\ref{fig:images}d because it is sensitive to other aspects of the input velocity field than the pure 
vorticity\footnote{It should be noted that there is subtle difference between a circular motion and a true vortex. 
A small gas parcel might be advected with the gas flow in a circular trajectory but its vorticity is only non-zero if the gas parcel is rotating in itself in addition to following the circular flow. 
Nevertheless the circular trajectory of a gas parcel might be still part of a larger vortical flow field and should thus be detected by the search algorithm, even though both the vorticity and vorticity strength may have small values at that particular position..  
%
%
The determination of vorticity and thus the detection of vortex flows is therefore ultimately limited by the spatial resolution and accuracy of the input velocity field. 
}

In the following, we will present tests for both the enhanced vorticity and vorticity strength method. 
In both cases, one snapshot after another is processed sequentially. 
Potential vortex candidates are then found from the local maxima in the input map in the current snapshot.
Next, a Gaussian kernel is fitted to the map around each vortex candidate, although a more sophisticated kernel function could be used in the future.
The fitted Gaussian is then subtracted from the original map. 
A new iteration follows in which the resulting map is again searched for local maxima, the new vortex candidates are fitted and the fitted Gaussians are again subtracted.
This iterative process is repeated until the peak falls below a threshold value of $10$\,\% of the maximum peak in the original map.
This procedure is similar to the CLEAN algorithm as used for interferometric imaging \citep{hogbom:1974} for the purpose of excluding irregular features and artefacts.
The size of a vortex is determined approximately as the diameter of a circle whose area is equivalent to the area enclosed by a contour line for intensity values larger than $10$\,\% of the intensity peak value.

\begin{figure}[tp!]
\centering
\includegraphics[width=1.0\columnwidth]{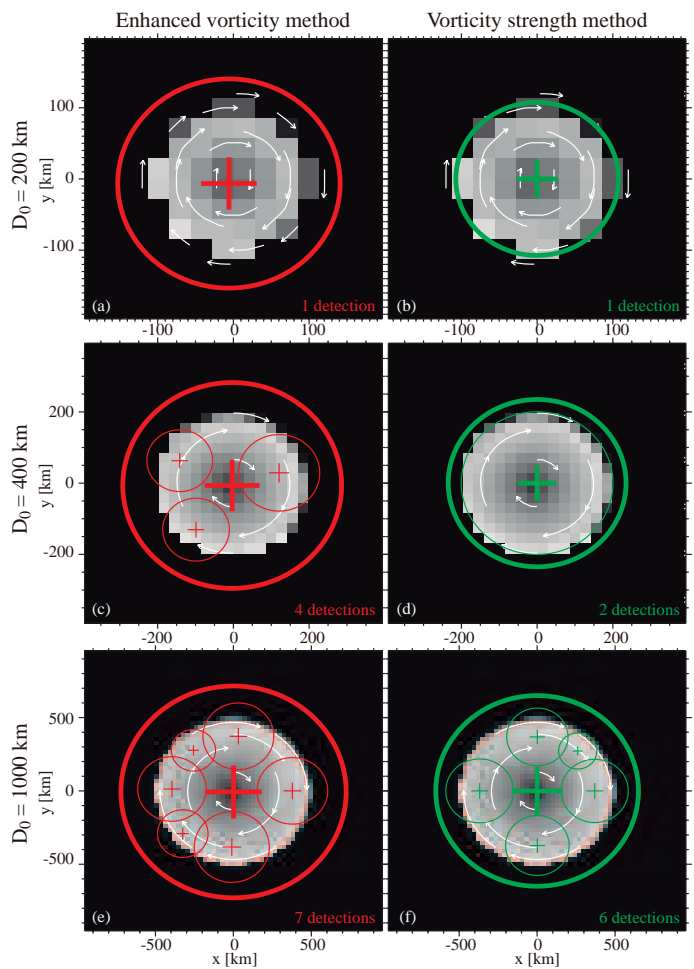}
\caption{Examples of the detection procedure for a vortex with a solid body rotation by using the enhanced vorticity method (left column) and the vorticity strength method (right column).  The diameters of the rotating feature $D_{\rm 0}$ is $200\,{\rm km}$ for (a) and (b),  $D_{\rm 0}=400\,{\rm km}$ for (c) and (d), and $D_{\rm 0}=1000\,{\rm km}$ for (e) and (f), respectively.  LIC imagings are shown in gray-scale in the background.  Red and green crosses and circles indicate the location and the size of detected vortices, respectively.}
\label{fig:compare}%
\end{figure}

The detections from a run based on the enhanced vorticity and another run based on the vorticity strength are compared in order to determine which of the quantities produces the most reliable and complete results. 
Figure~\ref{fig:sample} shows the difference between the LIC image, the vorticity, and the vorticity strength for a velocity field given by a solid rotation (upper panels) and by a shear flow (lower panels).
The intensity of the LIC image is proportional to the horizontal velocity amplitude showing a hollow structure, which is higher in the outer parts of the tested flow patterns (panels (a) and (d)). 
Distinguishing between vortex flow and shear flow thus requires to take into account the entire flow structure. 
This requirement is problematic for a more realistic velocity field (see Fig.~\ref{fig:images}a), where the boundaries between individual vortex flows can be difficult to see. 
Uncertainties in the velocity map derived from observations could severely limit the detection of vortices when using LIC imaging only. 
Both the vorticity and vorticity strength have a much stronger signature, which thus results in a more robust detection method. 
The vorticity has large values for tangential discontinuities as found at the boundary between a rotational flow and the background as shown in Fig.~\ref{fig:sample}b. 
Therefore, areas of high vorticity are outlining the boundary (and thus the area) of a vortex flow rather than the location of the centre of a vortex flow.
The vorticity strength, on the other hand, has the advantage that it can clearly discriminate between a solid rotation and a shear flow as can seen from comparing panels~(c) and~(f) in Fig.~\ref{fig:sample}.
Fig.~\ref{fig:compare} shows how our methods detect a swirling feature with different diameters of, $D_{\rm 0}=200\, {\rm km}$ for panels (a) and (b),  $D_{\rm 0}=400\, {\rm km}$ for (c) and (d),  and $D_{\rm 0}=1000\, {\rm km}$ for (e) and (f).
The enhanced vorticity method and the vorticity strength method both successfully detect the smallest swirling feature in panels~(a) and~(b) but return significantly different sizes as indicated by red and green circles.
The size of the isolated swirling feature measured by the enhanced vorticity method is larger than its actual size whereas the vorticity strength method produces more accurate results. 
The main reason for this difference is that the vorticity is sensitive to the tangential discontinuity at the edge of a swirling feature as already shown in Fig.~\ref{fig:sample}b while the vorticity strength is not sensitive to the discontinuity. 
%
%
For the larger swirling features in \mbox{panels~c-f}, on the other hand, both methods tend to detect multiple features within the same vortex flows.
The largest circles mark the sizes for the detections in the first iteration. 
The sizes measured by both methods are similar but larger than the actual size of the vortex flow.
The multiple detections, which appear as smaller circles, are found in further iterations due to residues remaining after subtracting a fitted feature with a Gaussian kernel from the actual feature, which has a sharp edge. 
The areas of the multiple small-size detections often overlap and can be combined into one detection, which correctly represents the given vortex flow. 
In this particular simple and clear test case, the small vortex candidates even coincide with the large candidate detected in the first iteration and are therefore unnecessary.  
These unwanted, artificial detections are removed by a  post-processing in each frame.

\subsection{Event identification}
\label{subsec:event_identification}

\begin{figure*}
\centering
\includegraphics[width=0.9\hsize]{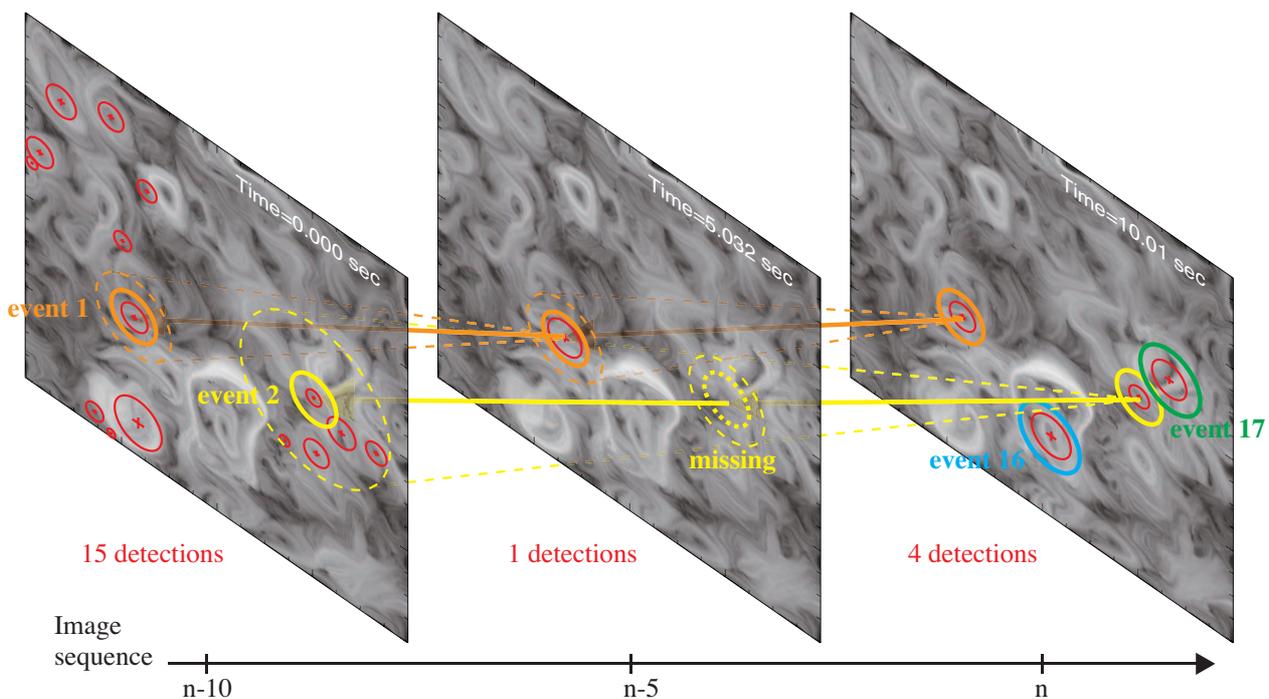}
\caption{Schematic illustration of how  the detected features (red crosses in circles) are attributed to the same events in an image sequence.   Orange, yellow, light-blue, and green circles indicate the identified events (\#1, 2, 16, and 17).  The orange and yellow dashed lines illustrate the characteristic cone corresponding to a fixed characteristic speed (e.g., either the sound speed or the Aflv\'en speed).   We choose to scan $10$ timesteps backwards in order to check the connectivity between the detected features and to confirm that they belong to the same event as illustrated for event\,2.  Involving several timesteps is important in case that the event is not properly or not all detected in an intermediate timestep. 
In case of multiple detection candidates for the connectivity (e.g., multiple detections within a dashed yellow circle), we select the candidate being closest in terms of location and size.} 
\label{fig:events}%
\end{figure*}

The previous steps return a number of consolidated vortex detections for each snapshot of the input image sequence. 
In this last step, detections from consecutive snapshots are combined into actual vortex events. 
The procedure is illustrated in Fig.~\ref{fig:events}. 
In order to decide if detected features belong to the same event or not, the geometrical distance between the features in the different snapshots must be less than the distance according to a set maximum speed and the time difference between the analysed snapshots.  
Here, the maximum speed is set to $50\,{\rm km/s}$ which is the maximum Alfv\'en speed occurring in the simulation.
Any larger distance would then implied that the difference in position of two detections is not physical and that these detections are not connected.  
We choose to scan 10~timesteps backwards (approximately $10$ seconds in this case) in order to check connectivity of the detected features. 
In cases of multiple candidates within the limiting distance, the detection is chosen that leads to the smallest change in size between the two detected features. 
The procedure could be replaced with a more sophisticated method using the cross-correlation function in a future upgrade.

\subsection{Tests with more realistic flow fields}
\label{subsec:marker_tests}

In the previous sections, we have shown that both the enhanced vorticity method and the vorticity strength method can detect an isolated vortex flow with a solid body rotation. 
However, typical flow fields in the solar atmosphere are much more complicated and the successful detection of vortex flows is much more challenging task. 
In order to demonstrate how our method works for more realistic flow fields in the solar atmosphere, we add a steady vortex with a solid body rotation to our realistic model atmosphere.
This combination has the advantage that the properties of the vortex flow are well known but that the fluctuations of the background are more realistic. 

\begin{table*}
\centering          
\begin{tabular}{l l c c c c c c }
\hline\hline       
 & & \multicolumn{3}{c}{Enhanced vorticity method} & \multicolumn{3}{c}{Vorticity strength method}\\
\multicolumn{2}{r}{$D_\mathrm{0}$}                         & $200$\,km & $400$\,km & $1000$\,km & $200$\,km & $400$\,km & $1000$\,km\\ 
\hline
   \multicolumn{2}{l}{Number of detected events}     & $1714$ & $1659$  &  $955$  & $1802$ & $1205$ & $828$\\
   \multicolumn{2}{l}{Detection rates\,[\%]}                & $86.9$  & $72.4$   &  $95.0$ & $99.9$  & $99.9$  & $99.9$\\ 
   Vortex position&- Overall accuracy\,[\%]               & $77.5$  & $53.6$   &  $20.2$  & $96.0$ & $95.5$  & $98.4$\\
                              &- Mean offset ($\bar{d}$)\,[km] & $22.5$  & $92.8$   & $399.0$ & $4.0$   & $8.9$    & $8.0$\\
   Vortex diameter&- Overall accuracy\,[\%]              & $69.7$  & $49.8$  &  $31.2$   & $78.2$ & $65.0$   & $62.6$\\
                  &- Mean error ($\overline{\Delta D}$)\,[km]    & $40.4$  & $2.0$    & $-554.3$ & $33.3$ & $117.0$ & $366.1$\\
\hline                  
\end{tabular}
\caption{Performance results for the enhanced vorticity and the vorticity strength method based on tests with artificial vortex flows superimposed on a ``realistic'' simulated flow field.
The following results are compared for both methods for prescribed vortex diameters of 200\,km, 400\,km, and 1000\,km: 
The number of detected vortices, 
the detection rate~(Eq.~(\ref{eqn:detection_rate})), 
the position accuracy~(Eq.~(\ref{eqn:position_accuracy})), 
the corresponding mean position offset~(Eq.~(\ref{eqn:mean_distance})), 
the accuracy of determined vortex diameters~(Eq.~(\ref{eqn:size_accuracy})), 
and the mean error in returned vortex diameter~(Eq.~(\ref{eqn:size_error})).
}
\label{table:diagnosis}      
\end{table*}

\begin{figure}[ht]
\centering
\includegraphics[width=0.9\hsize]{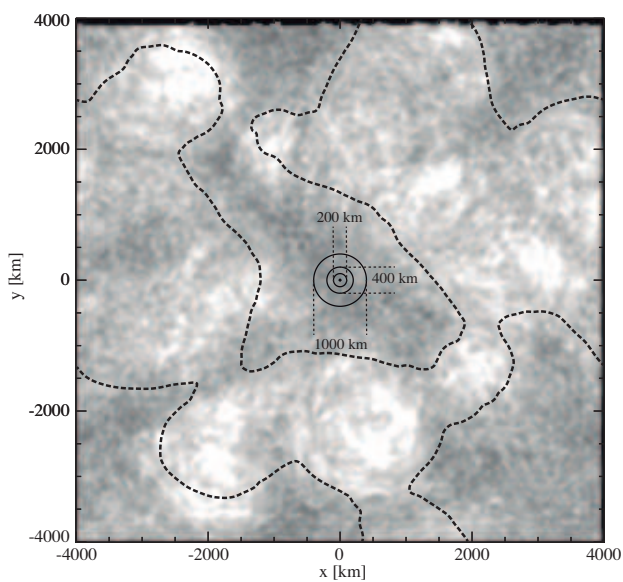}
\caption{LIC imaging averaged over the sampled time period of $30$ minutes in the simulation.  The location of the test vortices are indicated by a black cross at ($x$,$y$)=($0$,$0$) and black circles with diameters of $D_{\rm 0}=200\,{\rm km}$, $400\,{\rm km}$ and $1000\,{\rm km}$, respectively.  The strong magnetic field region is indicated by a dashed contour line enclosing the region in which the time-averaged magnetic field strength is larger than the initial magnetic field strength of $50$\,G.  The region occupies $56$~\% of the area of the entire computational domain.}
\label{fig:marker}%
\end{figure}

Figure~\ref{fig:marker} shows the time-average of the LIC images over the time period of $30$ minutes.
The time-averaged LIC image corresponds well to the magnetic field distribution as  indicated by a dashed contour line enclosing regions where the time-averaged magnetic field strength is larger than the initial magnetic field strength of $50\,{\rm G}$.
A steady vortex flow is added as a reference case in the central region at $(x,y)=(0,0)$, where there are relatively few significant features otherwise.
As in the simplified test case used before, a vortex flow with solid body rotation and a fixed diameter $D_\mathrm{0}$ is used as reference case.   
Three different diameters of $D_\mathrm{0}=200$, $400$, and $1000\,{\rm km}$ are tested for the reference vortex. 
The detection of the added vortex flow may depend on the interference from the nearby swirling features and the background, which may affect the resulting lifetime and the size of the artificial test vortex. 

As a measure for the success of the detection method, we estimate the detection rate of the test vortex as ratio of the number of frames with a successful detection, $n$, and the total number of frames during the simulation time of $T=30\,{\rm minutes}$, $N$:
\begin{equation}
\hbox{Detection rate\,(\%)} = \frac{n}{N}\times 100\,.
\label{eqn:detection_rate}
\end{equation}
We also measure the offset distance $d_{\rm i}$ of the detected test vortex from the original position in each frame, $i$, and obtain the mean offset distance $\bar{d}$:
\begin{equation}
\bar{d} = \frac{1}{n}\sum\limits_{i=1}^{n}{d_{\rm i}},
\label{eqn:mean_distance}
\end{equation}
and evaluate the position accuracy as the ratio between the mean offset distance $\bar{d}$ and the original radius of the test vortex~\mbox{$r_0 \equiv D_0/2$}:
\begin{equation}
\hbox{Position accuracy\,(\%)} = \left(1 - \frac{\bar{d}}{r_{\rm 0}}\right)\times 100\,.
\label{eqn:position_accuracy}
\end{equation}
The deviation $\sigma$ of the diameter of a test vortex $D_\mathrm{i}$ in each frame as derived with the detection algorithm from the original diameter of the test vortex $D_0$ is given by
\begin{equation}
\sigma = \frac{1}{n}\sqrt{\sum\limits_{i=1}^{n} (\Delta D_\mathrm{i})^{2}} \hbox{ with } \Delta D_\mathrm{i} = D_\mathrm{i} - D_0 \,,
\label{eqn:size_deviation}
\end{equation}

and the mean error $\overline{\Delta D}$ in returned vortex diameter is 
\begin{equation}
\overline{\Delta D}\equiv \frac{1}{n}\sum\limits_{i=1}^{n}{\Delta D_\mathrm{i}}\,.
\label{eqn:size_error}
\end{equation}
The relative accuracy in detected size can then be expressed as 
\begin{equation}
\hbox{Vortex diameter accuracy\,(\%)} = \left(1 - \frac{\sigma}{D_0}\right)\times 100
\label{eqn:size_accuracy}
\end{equation}
with a size ratio of 100\,\% representing a perfect result. 
The results are summarized in Table\,\ref{table:diagnosis}. 
%
\begin{figure}[t]
\centering
\includegraphics[width=1.0\hsize]{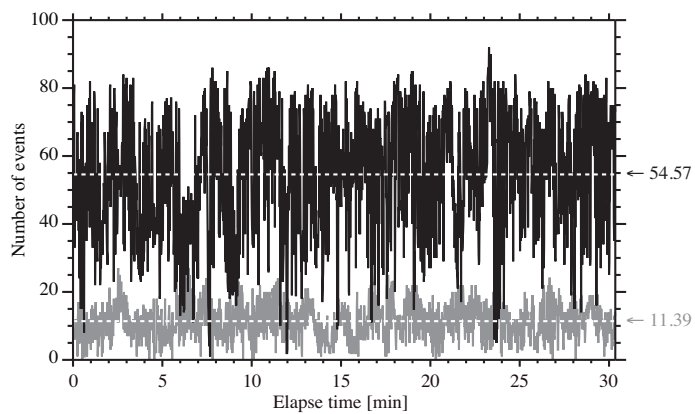}
\caption{Number of events detected with the vorticity strength method (grey curve) and with the enhanced vorticity method (black curve).  The time-averaged numbers of events detected with the vorticity strength and the enhanced vorticity methods are $55$ and $11$, respectively.}
\label{fig:numbers}%
\end{figure}

Based on this test, the vorticity strength method turns out to be superior to the enhanced vorticity method. 
The detection rate of the vorticity strength method is $99.9$\,\% independent of the size of the test vortex and thus higher than the corresponding rate for the enhanced vorticity method, which varies with vortex size but is 95.0\,\% at best (see Table~\ref{table:diagnosis}). 
The position accuracy of the vorticity strength method is more than $95$\,\% but varies slightly, depending on the size of the test vortex. 
At the same time, the method always overestimates the actual size of the test vortex, resulting in a size accuracy between 60\,\% and 80\,\%. 
These results are nevertheless better than for the enhanced vorticity method as described below. 
The enhanced vorticity method has a smaller rate of successful detections ranging from $70$\,\% to $95$\,\%, depending on the size of the to-be-detected test vortex. 
Furthermore, the positions and sizes returned by the enhanced vorticity method are less accurate than for the vorticity strength method. 
It is important to emphasise that the test cases in Fig.~\ref{fig:compare} have very sharp edges, whereas the transition between the vortices and the background flow field in the more realistic test case is much smoother. 
The enhanced vorticity method contains the convolution of streamlines and vorticity maps, which effectively results in the smoothing of gradients.
While this degrading effect still leaves clear enough boundaries in the test case with isolated vortices, these boundaries become to unclear for the test case with a more realistic backgrounds, resulting in low performance of the enhanced vorticity method. 
The vorticity strength approach, on the other hand, is able to detect the actual size of the swirling features without any problems because the method does not involve convolution with streamlines and thus no degrading of the vorticity strength maps. 
The method therefore performs much better for both types of test cases. 
%

\section{Results}
\label{sec:results}

In the following, we present the results for all vortex flows detected in 30\,min long simulation sequence 
(see Sect.~\ref{subsec:model}).

\subsection{\emph{Number of events}}
\label{subsec:numbers}

The total numbers of events found with the enhanced vorticity method and the vorticity strength method are $1720$ and $1600$, and hence the occurrence rates of events are $8.9\times 10^{-1}$~vortices Mm$^{-2}$ min$^{-1}$ and $8.3\times 10^{-1}$~vortices Mm$^{-2}$ min$^{-1}$, respectively. 
While these numbers are quite similar, the number of events in each time step (i.e. the instantaneous vortex detections) are quite different. 
The difference becomes obvious when plotting the number of events as a function of time (see Fig.~\ref{fig:numbers}\,a).
For both methods, the numbers vary strongly in time. 
On average, the vorticity strength method detects $55$~vortices in each time step compared to only $11$~vortices per time step with the enhanced vorticity method.
The enhanced vorticity method detects fewer events with shorter lifetime but then combines a larger number of them into the same events as compared to the vorticity strength method, resulting in comparable overall numbers of events. 
%

\subsection{Lifetime of swirling features}
\label{subsec:lifetime}

\begin{figure*}
\centering
\includegraphics[width=0.9\hsize]{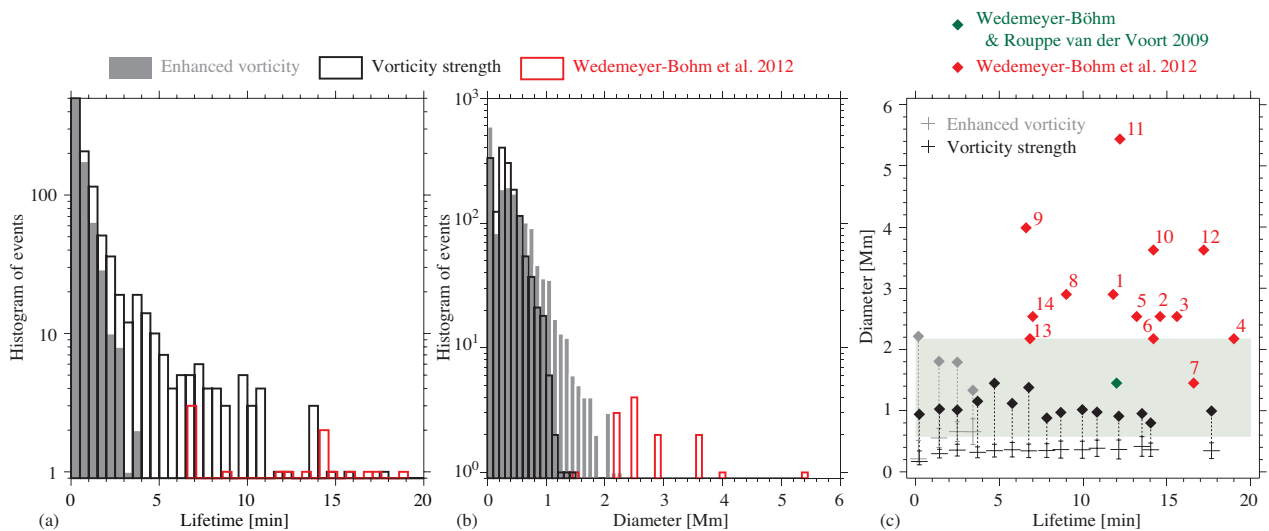}
\caption{Histograms of (a)~the lifetime of the detected events and (b)~the maximum diameter of the detected events. 
In panel~(c), the lifetime is plotted versus vortex diameter.
In panels~a and b, the results found with the enhanced vorticity method are plotted with grey bars, whereas results of the vorticity strength method are shown as black lines.  
For comparison, the chromospheric swirls found by  \citet{wedemeyer-bohm+:2012} are plotted with red lines.
In panel~c, all events detected in the simulation are grouped according to their lifetime in bins of 1\,min. 
For each group, the average value (horizontal line) plus/minus one standard deviation (shorter horizontal line above and below) and the maximum value (diamond) are plotted. 
Again, the results of the enhanced vorticity method are plotted in grey, whereas the results of the vorticity strength method are shown in black.  
For comparison, the chromospheric swirls observed by  \citet{wedemeyer-bohm+:2012} are plotted as red diamonds with numbers according to the table in the supplementary information of their study. 
In addition, the clearest swirl described by \citet{wedemeyer-bohm+rouppe_van_der_voort:2009} is represented as a green diamond, whereas the range of swirl diameters found in that study is marked as a green shaded area. 
}
\label{fig:histogram}%
\end{figure*}

%
In Fig.~\ref{fig:histogram}\,a, histograms for event lifetimes are shown as derived with the enhanced vorticity method (gray shade) and the vorticity strength method (black).
For the enhanced vorticity method, the mean lifetime of the events is $16.0$\,s with a the standard deviation of $26.5$\,s, whereas the vorticity strength method produces events with a mean lifetime of $52.0$\,s and a standard deviation of $113$\,s.
The events detected by the enhanced vorticity method last not more than $5$\,min, whereas the vorticity strength method results in a significant number of events with lifetimes of more than that.
There is a particular event found by the vorticity strength method, which seems to last $22$\,min.
This particular case is due to clusters of swirling features, which appear very close to each other. 
It is therefore not straightforward, in particular for an automated detection algorithm, to decide if an event persists as a single entity or if it is rather an intricate succession of splitting and merging vortices.
Furthermore, this rather extreme lifetime exceeds the so far longest reported lifetime for a chromospheric swirl of 19\,min in the sample presented by \citet{wedemeyer-bohm+:2012}. 
Such extreme events must thus be considered with some caution but such events are also quite rare. 
It might be possible to further split such events by using a cross-correlation function into sub-events but more detailed investigations leading to a more precise definition of vortex events are needed. 
%

\subsection{Size of swirling features}
\label{subsec:size}

%
Fig.~\ref{fig:histogram}b shows histograms of the maximum diameter of vortex events detected by using both the enhanced vorticity method (gray shade) and the vorticity strength method (black).
There are significant differences between them.
The enhanced vorticity method results in a mean diameter of  533\,km with a standard deviation of 274\,km, whereas the vorticity strength method produces a mean radius of $338$\,km with a standard deviation of 132\,km.
As we described in Sect.\,\ref{subsec:marker_tests}, the effective diameters measured by the enhanced vorticity method tend to be larger than those measured by the vorticity strength method.
The number of events detected by the enhanced vorticity method drops sharply for diameters $> 2000$\,km.
The maximum diameter of the events detected by the enhanced vorticity method is  2215\,km whereas that detected by the vorticity strength is  1449\,km.
While the chromospheric swirls observed by \citet{wedemeyer-bohm+:2012} have diameters on average of $(2.9 \pm 1.0)$\,km with the smallest value being 1.5\,Mm, almost all events detected in our study are smaller than $2$~Mm.

\subsection{Correlation between lifetime and size}
\label{subsec:correlation}

%
In Fig.~\ref{fig:histogram}c, the joint probability distribution for the lifetime and the size of detected events is compared to the observations of chromospheric swirls by \citet{wedemeyer-bohm+:2012}. 
It seems that the sizes of the  events detected in our simulation are distinctively different from the those of the observed  chromospheric swirls as we already mentioned in Sect.\,\ref{subsec:size}.
Furthermore, there is no obvious relation between size and lifetime based on the current simulation-based sample. 
There is nevertheless some indication that the lifetimes and sizes correlate. 
This finding needs a more systematic analysis and will be addressed in a forthcoming publication.

\section{Discussion}
\label{sec:discussion}

\subsection{Occurrence rate of vortex flows}
\label{subsec:occurrence_rate}
\begin{figure}
\centering
\includegraphics[width=1.0\hsize]{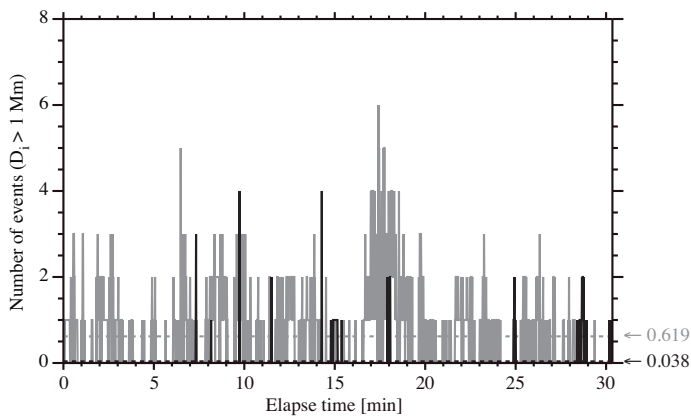}
\caption{Same as Fig.~\ref{fig:numbers} but for the events whose diameter is larger than $1000$ km.}
\label{fig:numbers500}%
\end{figure}

The occurrence rate of vortices is a quantitative indicator of the complexity of flows in the solar atmosphere. 
The event occurrence rates produced by the enhanced vorticity method and the vorticity strength method are both close to a value of $1$ vortex Mm$^{-2}$ minute$^{-1}$ (see Sect.~\ref{subsec:numbers}), which implies that a chromospheric vortex flow could exist for every granular cell. 
On the other hand, the value is four orders of magnitude larger than the value derived from observations of chromospheric swirls \citep[e.g.,][]{wedemeyer-bohm+:2012} and three orders of magnitude larger than the value derived from observations of photospheric vortices \citep[e.g.,][]{bonet+:2010}. 
This apparent discrepancy can be explained to large extent with the limitations of observations as compared to the simulation  analysed here. 
As can be seen in Fig.~\ref{fig:histogram}, there is a large number of small-scale and short-lived chromospheric vortex flows in the simulation. 
Those events have a mean lifetime in the range of $10 - 200$ seconds (Sect.\,\ref{subsec:lifetime}) and a mean diameter of  roughly $400$\,km corresponding to about $\sim 0.5$\,arcsec. 
In comparison, \citet{wedemeyer-bohm+rouppe_van_der_voort:2009} observed events with radii down to 0.4-0.6\,arcsec and thus diameters on the order of 600-900\,km, although a reliable analysis of such small events is already at the limit and most of their results are based on larger examples with diameters of up to $\sim 2200$\,km (see shaded area in Fig.~\ref{fig:histogram}c). 
%

\subsection{Observational limitations}
\label{subsec:observ_limit}

The successful detection of a vortex flow in an observational image sequence is a challenging task, which can be summarized with the following requirements:  
\begin{itemize} 
\item The sub-structure of the vortex swirl, e.g., a ring, must be resolved clearly. 
\citet{wedemeyer-bohm+rouppe_van_der_voort:2009} report ring widths of only $0.2$\,arcsec, which is close to the angular resolution that can typically be achieved with current solar telescopes. 
The width of the rotating ring described by \citet{wedemeyer-bohm+rouppe_van_der_voort:2009} is about $10$\,\% of the swirl diameter. 
Scaling these proportions to a vortex diameter of only $\sim 0.5$\,arcsec would imply ring widths of only 0.05\,arcsec, which clearly beyond what can be currently resolved. 
\item The cadence of observation must be high enough so that the plasma motions can be tracked reliably. 
\citet{wedemeyer-bohm+rouppe_van_der_voort:2009} estimate horizontal speeds of the order of $10$\,km/s. 
A feature rotating as part of a chromospheric vortex with that speed would move about $0.1$\,arcsec within roughly $7$\,sec. 
A temporal resolution on that order (or better) is thus needed. 

\item A rotating motion should be visible in a number of consequent frames. 
Chromospheric observations often have cadences of $10$\,s or more.
Detecting vortex flows with lifetimes of only a few $10$\,s is thus at the limit, in particular because rotating motions can easily be obscured by the generally complex dynamic pattern of the chromosphere.
Reliable detections would thus favour longer-lived vortex events with at least one clear rotation cycle. 

\item Variations in seeing conditions may make some frames less clear and thus reducing the overall effective cadence of the image sequence and thus the detectability of a vortex flow. 

\end{itemize}

These requirements result naturally in limits for the shortest lifetimes and smallest diameters of vortex flows that can reliably be observed. 
The abundant small-scale events found in this paper would therefore remain undetected with current solar telescopes but pose an interesting science use case for the next generation of solar telescopes such as Daniel K. Inouye Solar Telescope \citep[DKIST, formerly the Advanced Technology Solar Telescope, ATST:][]{2011ASPC..437..319K} and the 4 m European Solar Telescope \citep[EST:][]{2010AN....331..615C}
\subsection{Large-scale vortex flows} 
\label{subsec:lifetime_largescale}

For the comparison with observations, we now select all detected events with diameters larger than $1000$\,km. 
The number of large-scale events is plotted as a function of simulation time in Fig.~\ref{fig:numbers500}. 
The occurrence rates of large-scale events found with the enhanced vorticity method and the vorticity strength method are  $7.1\times 10^{-2}$~vortices Mm$^{-2}$ minute$^{-1}$ and $8.8\times 10^{-3}$~vortices Mm$^{-2}$ minute$^{-1}$, respectively. 
The latter is similar to the occurrence rate of observed photospheric vortices. 
This finding supports the conclusion that the discrepancy between the simulation results presented here and previous observations may to large extent result from undetected small-scale events due to the lack of spatial/temporal resolution in  observations. 
For the large-scale events (i.e. with diameters $> 1000$\,km) found with the enhanced vorticity method, the mean lifetime is $42.7$~s and the maximum lifetime is $214$~s. 
For those detected with the vorticity strength method, the maximum lifetime is $22$~min and the mean lifetime is $272$~s 
as compared to 52\,s for the whole sample (see Sect.\,\ref{subsec:lifetime}). 
This finding implies that large-scale events tend to have relatively longer lifetime although this trend is not very obvious in Fig.~\ref{fig:histogram}c. 
The diameter for the large-scale events produced with the enhanced vorticity method become larger than $1000$~km continuously during their entire lifetime, however, the diameter of those with the vorticity strength method becomes larger than $1000$~km during only a fraction of their lifetime. 
Interestingly, for both methods, the periods with diameter being larger than $1000$~km last typically no more than $60$~s. 
%

\subsection{Role of magnetic fields}
\label{subsec:magnetfield}

Photospheric vortex flows seem to occur preferably within intergranular lanes and even more so at lane vertices. 
While photospheric vortex flows form as natural ingredient of surface convection flows, the formation of a chromospheric vortex flow usually requires in addition that the photospheric footpoint of a magnetic field structure coincides with a photospheric vortex flow. 
This way, rotating motions in the photosphere can be transferred into the chromosphere \citep{wedemeyer+steiner:2014}. 
Consequently, not all photospheric vortex flows have a corresponding chromospheric vortex and the occurrence rate of photospheric events is higher than the corresponding chromospheric one as supported by observations 
(see Sect.~\ref{sec:intro}). 
The simulations presented here again confirm that finding, namely that the locations of chromospheric vortex flows are clearly connected to the magnetic field topology. 
For instance, the intensity of the LIC image in Fig.\,\ref{fig:marker} is an indicator for vortex motions and shows a clear correspondence to regions with stronger magnetic fields (dashed contour line in the figure). 
It thus safe to conclude that at least the majority of chromospheric vortex flows requires the presence of magnetic fields and that the vortex properties will depend on the properties of the magnetic field. 
%

\subsection{Size of vortex flows} 
\label{subsec:size_limit}

The maximum diameter of chromospheric vortex flows detected in our study is $\sim 2.1$~Mm (see Fig.\,\ref{fig:histogram}b), which corresponds to 1-2 granules in the photosphere.
The size seems to be consistent with the occurrence rate of $1$ vortex Mm$^{-2}$ minute$^{-1}$  (Sect.\,\ref{subsec:occurrence_rate}), which seems plausible in view of the strong coupling of chromospheric vortex flows and their photospheric counterparts as mediated by magnetic fields (see Sect.~\ref{subsec:magnetfield}). 

The maximum diameter of vortex flows found in the simulation presented here is smaller than most events found in observations. 
Next to reasons connected to the aforementioned photospheric-chromospheric coupling, the size of the computational box and the properties of the modelled magnetic field clearly affect the maximum vortex size. 
Our simulation box has an extent of only $8$~Mm in both horizontal directions and also has periodic horizontal boundary conditions.  
The magnetic field, which was initially purely vertically aligned, was rearranged into magnetic flux concentrations, which are mostly rooted in the intergranular lanes in the photosphere. 
The fraction of horizontal area with magnetic field strengths in excess of $50$\,G is $56$~\% (see the contour in Fig.\,\ref{fig:marker}). 
Let us consider that there is a magnetic flux concentration located at every photospheric intergranular vertex, that all flux concentrations have the same polarity, and that all of them expand with height in an idealized wineglass shape. 
In this case, the extent of any magnetic field structure in the chromosphere would be effectively limited by the field of the neighbouring, also expanding field structures. 
Consequently, the granular spatial scale of $1 - 2$\,Mm in the photosphere is also imprinted on the chromospheric magnetic field. 
In a less idealized setting with a more complex magnetic field topology, the details certainly depend on the actual coverage of the strong magnetic field regions in the simulation box but the maximum vortex diameters still seem reasonable in view of the box dimensions. 
Simulations of chromospheric vortex flows with diameters similar to what has been observed so far would require correspondingly larger computational boxes. 
On the other hand, the simulations by \citet{rempel:2014} exhibit peaks of both kinetic and magnetic energy at a scale of $2$~Mm in the photosphere independent of the simulation box size in the range from $6.1$~Mm $\times 6.1$~Mm to $24.6$~Mm $\times 24.6$~Mm.
Given the photospheric-chromospheric coupling as discussed in Sect.~\ref{subsec:magnetfield}, it seems likely that this characteristic scale of $2$~Mm in the photospheric magnetic field would also affect the sizes of chromospheric vortex flows. 
However, observations clearly show that vortex flows with larger diameters occur. 
For instance, \citet{1988Natur.335..238B} found a photospheric vortex with a diameter of 5\,Mm, which thus spans over a few granules.  
By means of local correlation tracking, \citet{2016arXiv161007622R} study the horizontal flow field in photospheric observations and find indeed indications for meso-granular scales, i.e. spatial scales equivalent to a few granules. 
The radiative hydrodynamic simulations of \citet{lord+:2014}, which have a size of $192$~Mm $\times 196$~Mm box and which include the treatment of helium ionisation, exhibit a peak of kinetic energy in the upper convection at a scale of $\sim20$~Mm. 
To our knowledge, no chromospheric vortex flow with a diameter larger than $\sim6$\,Mm has been observed so far (see 
Fig.~\ref{fig:histogram}) but vortex flows or more generally rotating motions may exist of a very large range of spatial scales \citep[see, e.g.,][]{wedemeyer+:2013b}.

On the other end of the distribution,  small photospheric vortex flows are found in the simulations by 
\citet{moll+:2011,2012A&A...541A..68M} with an average lifetime of $3.5$~minutes, suggesting that vortex flows on a large range of scales are an integral part of dynamics driven by surface (magneto)convection. 
The photospheric-chromospheric coupling through magnetic fields implies that an equally large distribution of scales should also exist for vortex flows in the chromosphere, as suggested by our results  in Fig.~\ref{fig:histogram}c. 
%

\section{Conclusions}
\label{sec:conclusions}

In this first part of a series, we have introduced two automatic methods for detecting vortex flows and tested them on a 
3D numerical magnetohydrodynamic simulation of the solar atmosphere. 
We conclude that the vorticity strength method is superior in all aspects compared to the enhanced vorticity method. 
The vorticity strength method is more successful in detecting and locating swirling features and also gives more accurate vortex diameters.  
Nevertheless, the enhanced vorticity method can detect \emph{additional} events with rotating motions but low vorticity values, which would otherwise remain undetected. 
The consequences of the limited accuracy of velocity fields derived from local correlation tracking techniques, such as they would be derived from observational data sets, will be discussed in a forthcoming publication.

Immediate results from applying the detection methods on the 3D numerical simulation are:  
\begin{enumerate}
  \item Chromospheric vortex flows are very abundant in the analysed simulation. 
  \item There is a continuous distribution in vortex diameter and lifetime with the short-lived and smallest vortex flows being most abundant. 
  \item Many of the detected vortex flows would be too small to be found in observations. 
  \end{enumerate}
These results, which are based on the analysed simulation,  imply that small-scale vortex flows might be very abundant in the solar chromosphere although they could not yet be observed due to instrumental limitations. 
The contributions to the transport of energy in the solar atmosphere might be small for individual small-scale events but their large number could result in a combined contribution that should be significant and should thus be investigated in more detail. 
Future larger telescopes such as DKIST and EST would allow for observations of smaller vortex flows and thus help to shed to light on this integral part of chromospheric dynamics. 

\begin{acknowledgements}
YK and SW acknowledge support by the Research Council of Norway, grant 221767/F20.  YK and SW express sincere thanks to Oskar Steiner for providing support.\\
\end{acknowledgements}

%
%

\bibliographystyle{aa}
\bibliography{ms}

\begin{thebibliography}{24}
\expandafter\ifx\csname natexlab\endcsname\relax\def\natexlab#1{#1}\fi

\bibitem[{Bonet {et~al.}(2010)Bonet, M{\'a}rquez, S{\'a}nchez~Almeida,
  Palacios, Mart{\'\i}nez-Pillet, Solanki, del Toro~Iniesta, Domingo,
  Berkefeld, Schmidt, Gandorfer, Barthol, \& Kn{\"o}lker}]{bonet+:2010}
Bonet, J.~A., M{\'a}rquez, I., S{\'a}nchez~Almeida, J., {et~al.} 2010, The
  Astrophysical Journal Letters, 723, L139

\bibitem[{{Brandt} {et~al.}(1988){Brandt}, {Scharmer}, {Ferguson}, {Shine}, \&
  {Tarbell}}]{1988Natur.335..238B}
{Brandt}, P.~N., {Scharmer}, G.~B., {Ferguson}, S., {Shine}, R.~A., \&
  {Tarbell}, T.~D. 1988, \nat, 335, 238

\bibitem[{Cabral \& Leedom(1993)}]{cabral+:1993}
Cabral, B. \& Leedom, L.~C. 1993, in SIGGRAPH '93: Proceedings of the 20th
  annual conference on Computer graphics and interactive techniques (ACM
  Request Permissions)

\bibitem[{Chong {et~al.}(1990)Chong, Perry, \& Cantwell}]{chong+:1990}
Chong, M.~S., Perry, A.~E., \& Cantwell, B.~J. 1990, Physics of Fluids A (ISSN
  0899-8213), 2, 765

\bibitem[{{Collados} {et~al.}(2010){Collados}, {Bettonvil}, {Cavaller},
  {Ermolli}, {Gelly}, {P{\'e}rez}, {Socas-Navarro}, {Soltau}, {Volkmer}, \&
  {EST Team}}]{2010AN....331..615C}
{Collados}, M., {Bettonvil}, F., {Cavaller}, L., {et~al.} 2010, Astronomische
  Nachrichten, 331, 615

\bibitem[{Fisher \& Welsch(2008)}]{fisher+welsch:2008}
Fisher, G.~H. \& Welsch, B.~T. 2008, Subsurface and Atmospheric Influences on
  Solar Activity ASP Conference Series, 383, 373

\bibitem[{{Freytag} {et~al.}(2012){Freytag}, {Steffen}, {Ludwig},
  {Wedemeyer-B{\"o}hm}, {Schaffenberger}, \& {Steiner}}]{freytag+:2012}
{Freytag}, B., {Steffen}, M., {Ludwig}, H.-G., {et~al.} 2012, Journal of
  Computational Physics, 231, 919

\bibitem[{H{\"o}gbom(1974)}]{hogbom:1974}
H{\"o}gbom, J.~A. 1974, Astronomy and Astrophysics Supplement, 15, 417

\bibitem[{{Jeong} \& {Hussain}(1995)}]{1995JFM...285...69J}
{Jeong}, J. \& {Hussain}, F. 1995, Journal of Fluid Mechanics, 285, 69

\bibitem[{{Keil} {et~al.}(2011){Keil}, {Rimmele}, {Wagner}, {Elmore}, \& {ATST
  Team}}]{2011ASPC..437..319K}
{Keil}, S.~L., {Rimmele}, T.~R., {Wagner}, J., {Elmore}, D., \& {ATST Team}.
  2011, in Astronomical Society of the Pacific Conference Series, Vol. 437,
  Solar Polarization 6, ed. J.~R. {Kuhn}, D.~M. {Harrington}, H.~{Lin}, S.~V.
  {Berdyugina}, J.~{Trujillo-Bueno}, S.~L. {Keil}, \& T.~{Rimmele}, 319

\bibitem[{Lord {et~al.}(2014)Lord, Cameron, Rast, Rempel, \&
  Roudier}]{lord+:2014}
Lord, J.~W., Cameron, R.~H., Rast, M.~P., Rempel, M., \& Roudier, T. 2014, The
  Astrophysical Journal, 793, 24

\bibitem[{Moll {et~al.}(2011)Moll, Cameron, \& Sch{\"u}ssler}]{moll+:2011}
Moll, R., Cameron, R.~H., \& Sch{\"u}ssler, M. 2011, \aap, 533, A126

\bibitem[{{Moll} {et~al.}(2012){Moll}, {Cameron}, \&
  {Sch{\"u}ssler}}]{2012A&A...541A..68M}
{Moll}, R., {Cameron}, R.~H., \& {Sch{\"u}ssler}, M. 2012, \aap, 541, A68

\bibitem[{November(1986)}]{november:1986}
November, L.~J. 1986, Applied Optics, 25, 392

\bibitem[{November \& Simon(1988)}]{november+simon:1988}
November, L.~J. \& Simon, G.~W. 1988, \apj, 333, 427

\bibitem[{{Park} {et~al.}(2016){Park}, {Tsiropoula}, {Kontogiannis},
  {Tziotziou}, {Scullion}, \& {Doyle}}]{2016A&A...586A..25P}
{Park}, S.-H., {Tsiropoula}, G., {Kontogiannis}, I., {et~al.} 2016, \aap, 586,
  A25

\bibitem[{Rempel(2014)}]{rempel:2014}
Rempel, M. 2014, The Astrophysical Journal, 789, 132

\bibitem[{{Requerey} {et~al.}(2016){Requerey}, {Del Toro Iniesta}, {Bellot
  Rubio}, {Mart{\'{\i}}nez Pillet}, {Solanki}, \&
  {Schmidt}}]{2016arXiv161007622R}
{Requerey}, I.~S., {Del Toro Iniesta}, J.~C., {Bellot Rubio}, L.~R., {et~al.}
  2016, ArXiv e-prints [\eprint[arXiv]{1610.07622}]

\bibitem[{Steiner {et~al.}(2013)Steiner, Rajaguru, Vigeesh, Steffen,
  Schaffenberger, \& Freytag}]{steiner+:2013}
Steiner, O., Rajaguru, S.~P., Vigeesh, G., {et~al.} 2013, Memorie della Societa
  Astronomica Italiana Supplement, 24, 100

\bibitem[{Wedemeyer {et~al.}(2013)Wedemeyer, Scullion, Rouppe van~der Voort,
  Bosnjak, \& Antolin}]{wedemeyer+:2013b}
Wedemeyer, S., Scullion, E., Rouppe van~der Voort, L., Bosnjak, A., \& Antolin,
  P. 2013, The Astrophysical Journal, 774, 123

\bibitem[{Wedemeyer \& Steiner(2014)}]{wedemeyer+steiner:2014}
Wedemeyer, S. \& Steiner, O. 2014, Publications of the Astronomical Society of
  Japan, 66, 10

\bibitem[{Wedemeyer-B{\"o}hm \& Rouppe van~der
  Voort(2009)}]{wedemeyer-bohm+rouppe_van_der_voort:2009}
Wedemeyer-B{\"o}hm, S. \& Rouppe van~der Voort, L. 2009, \aap, 507, L9

\bibitem[{Wedemeyer-B{\"o}hm {et~al.}(2012)Wedemeyer-B{\"o}hm, Scullion,
  Steiner, Rouppe van~der Voort, de~La~Cruz~Rodriguez, Fedun, \&
  Erd{\'e}lyi}]{wedemeyer-bohm+:2012}
Wedemeyer-B{\"o}hm, S., Scullion, E., Steiner, O., {et~al.} 2012, Nature, 486,
  505

\bibitem[{Welsch {et~al.}(2004)Welsch, Fisher, Abbett, \&
  Regnier}]{welsch+:2004}
Welsch, B.~T., Fisher, G.~H., Abbett, W.~P., \& Regnier, S. 2004, \apj, 610,
  1148

\end{thebibliography}

\onecolumn
\ \\ 
\appendix
\section{Technical details}
\label{sec:appendix1}
\ \\ 
\begin{figure*}
\centering
\includegraphics[width=0.75\textwidth]{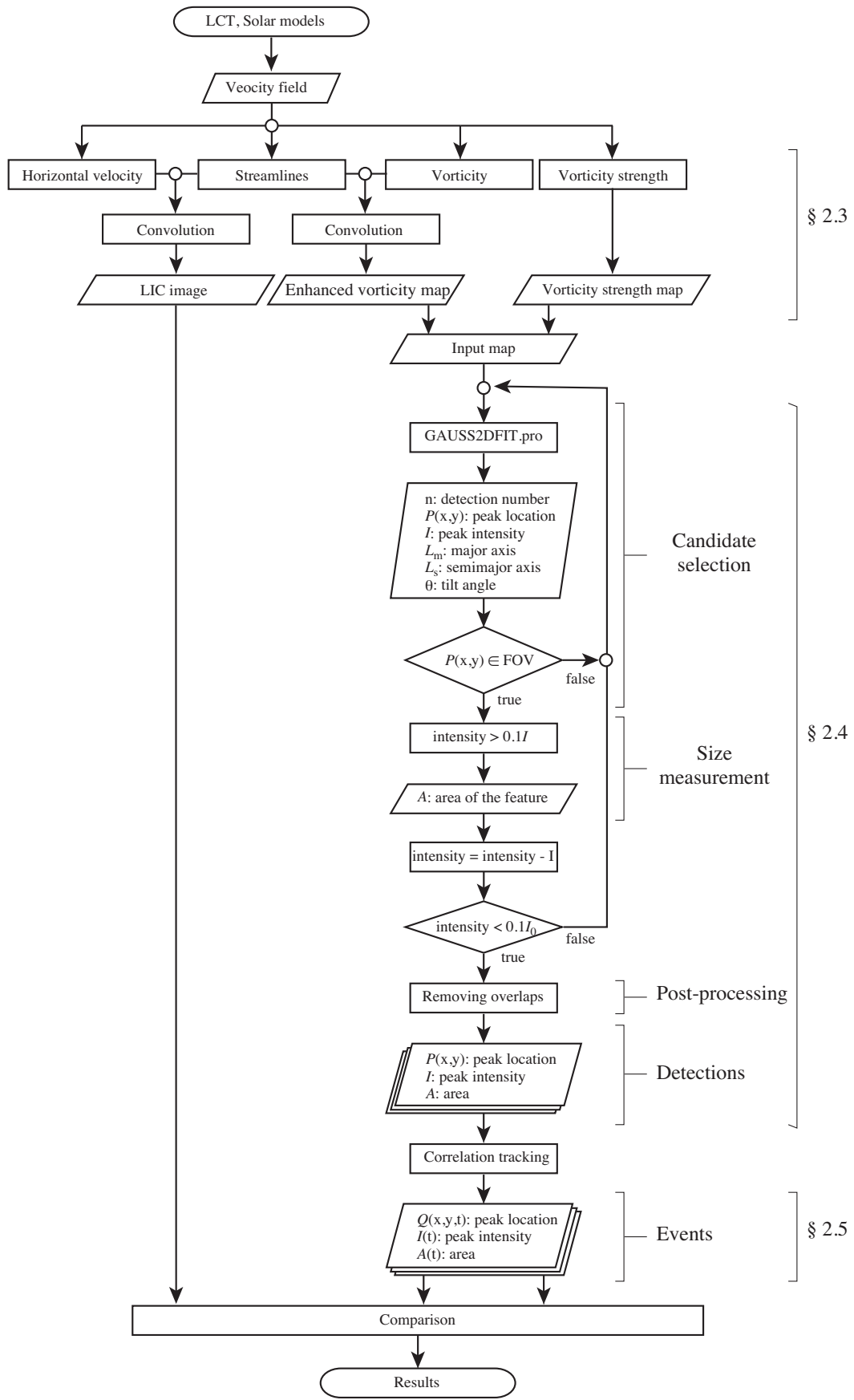}
\caption{Flowchart illustrating the procedure for detecting vortex flows with the enhanced vorticity method and the vorticity strength method.}
\label{fig:flowchart}%
\end{figure*}


\end{document}